\documentclass{mn2e}
\usepackage{graphicx}
\usepackage{amssymb, amsmath}

\def \bnabla {{\boldsymbol{\nabla}}}
\def \br {{\bf r}}
\def \bx {{\bf x}}

\def \bv {{\bf v}}
\def \bn {{\bf n}}
\def \bu {{\bf u}}

\begin{document}

\title{Why there is no Newtonian backreaction}
\author[Nick Kaiser]{Nick Kaiser\\
Institute for Astronomy, University of Hawaii, 2680 Woodlawn Drive, Honolulu, HI 96822-1839, USA}
\maketitle

\begin{abstract}
In the conventional framework for cosmological dynamics the scale
factor $a(t)$ is assumed to obey the `background' Friedmann equation 
for a perfectly homogeneous universe while particles move according
to equations of motions driven by the gravity of  the density
fluctuations. It has recently  been suggested that the emergence of structure
modifies the evolution of $a(t)$ viaNewtonian (or `kinematic') backreaction and that
this may avoid the need for dark energy. Here we point out  that the
conventional system of equations is exact in Newtonian gravity
and there is no approximation in the use of the homogeneous universe
equation for $a(t)$. The recently proposed modification of Racz et al.\ (2017) does
not reduce to Newtonian gravity in the limit of low velocities.  We discuss the relation
of this to the `generalised Friedmann equation' of Buchert
and Ehlers. These are quite different things; their formula describes
individual regions and is obtained under the restrictive
assumption that the matter behaves like a pressure-free fluid whereas
our result is exact for collisionless dynamics and is an auxiliary relation
appearing in the structure equations.  We use the symmetry of the
general velocity autocorrelation function to show how Buchert's $\cal Q$
tends very rapidly to zero for large volume and that this does
not simply arise `by construction' through the adoption of periodic boundary conditions
as has been claimed. We conclude that, to the extent that Newtonian
gravity accurately describes the low-$z$ universe, there is no backreaction
of structure on $a(t)$ and that the need for dark energy cannot be avoided in this way.
\end{abstract}


\section{Introduction}

It is usually assumed that the spacetime in an inhomogeneous
cosmology may be described by a metric which is that of a homogeneous
FRW model $ds^2 = - dt^2 + a(t)^2 dx^2$, where $x$
is conformal position, 
with additional very small `weak field' metric perturbations.
This does not require that the density perturbations be small;
only that the velocities associated with structures are small
compared to $c$.
The scale factor for the background is assumed to obey the Friedmann equation
\begin{equation}
\ddot a + \frac{4 \pi}{3} G {\overline \rho} a = 0
\label{eq:Friedmann}
\end{equation}
for a homogeneous background universe with density ${\overline \rho} \propto a^{-3}$,
and where dot denotes time derivative.
The density here may be augmented by additional terms $\rho + 3 P / c^2$ for
homogeneous dark energy or radiation backgrounds satisfying the
appropriate continuity equations.
The peculiar (i.e.\ non-Hubble) motions of non-relativistic particles such as dark matter
or galaxies obey the `structure evolution' equations
\begin{equation}
\dot \bv + H \bv = - \bnabla \phi / a
\label{eq:vdot}
\end{equation} 
where $\bv \equiv a \dot \bx$, $H = \dot a / a$ and 
the spatial derivative is with respect to comoving coordinates $\bx$,
and where $\phi$ is a solution of Poisson's equation 
sourced by the density perturbation, i.e.\  
\begin{equation}
\nabla^2 \phi = 4 \pi G (\rho - {\overline \rho})a^2.
\label{eq:poisson}
\end{equation}
This system of equations, which may also be obtained in  Newtonian cosmology (Peebles 1980), 
may be used to find the evolution of linearised perturbations and solved in N-body codes
to obtain non-linear structure.

Some, however, going back at least to Ellis (1984), have 
questioned the validity of this as (\ref{eq:Friedmann}) 
is derived assuming that the Universe is homogeneous, which is obviously not the case.
To address this, Buchert and Ehlers (1997),
modelling the matter as a Newtonian pressureless fluid (only a crude  approximation once
multi-streaming occurs, but valid in the linear and quasi-linear regime), have found that the
size $a \equiv V^{1/3}$ of a region of volume $V$ containing mass $M$ obeys
\begin{equation}
3 \ddot a/ a +4 \pi G M / a^3 = {\cal Q} .
\label{eq:BE97}
\end{equation} 
Here ${\cal Q} = 2\langle (\theta - \langle \theta \rangle)^2 \rangle / 3 + 2 \langle \omega^2 - \sigma^2 \rangle$ 
where $\theta$ is the volume expansion rate $\sigma^2$ and $\omega^2$ are the shear and rotations squared,
and $\langle \ldots \rangle$ denotes an average over the volume.
As with (\ref{eq:Friedmann}) this may be augmented by including a cosmological constant.  

This is highly suggestive.  The equation of motion (\ref{eq:BE97}) for the linear size $a$ is strikingly similar to (\ref{eq:Friedmann})
but has an extra term containing quantities that are second order in the 
velocity shear $d{\bf v}/ d{\bf r}$. The quantities being averaged in ${\cal Q}$ are, like the
individual terms in (\ref{eq:Friedmann}), generally of order the inverse squared dynamical time $G \rho$, 
so one might naively think this would be a strong effect.  
However, as Buchert and Ehlers point out, for large volumes the actual effect is less than this.
As regards the implications for cosmology, they say that
`the average motion may be {\em approximately\/} given by a Friedmann
model on a scale which is larger than the largest existing inhomogeneities.',
but they also argue that
`the ``conspiracy assumption'' that ${\cal Q} = 0$ [\ldots] must be
considered a strong restriction on generality'.

Equation (\ref{eq:BE97}) is the basis of Newtonian (or `kinematic')  backreaction'; the idea that there
is a modification of the expansion rate caused by the emergence of structure.  It has been
studied by Buchert, Kerscher \& Sicka 2000,
who make some interesting claims; explored in N-body simulations by Kazimierczak 2016;
and has been widely discussed in reviews of backreaction (e.g.\ Buchert \& R\"as\"anen, 2012).

In a similar vein, Racz et al.\ (2017) have proposed that the successes of the $\Lambda$CDM
concordance cosmology can be obtained without the need for dark energy.
They say `Cosmological N-body simulations integrate Newtonian dynamics 
with a changing GR metric that is calculated from averaged quantities.' but that
`There is a choice in how the averaging is done.'  
They propose to maintain equations (\ref{eq:vdot}) and (\ref{eq:poisson})
but obtain $a(t)$ by averaging the local expansion rate $\dot a / a$ computed from 
the local density under some simplifying assumptions and then using this
to update $a(t)$ at each time-step.   Performing N-body calculations
using this algorithm and with matter only they find $a(t)$
very similar to the solution of the Friedmann equation
in $\Lambda$CDM.  They argue that the successes of the concordance
cosmology can thereby be retained without the need for dark energy through this
`strong backreaction' effect.

But is it really legitimate to assume that backreaction from structure causes
$a(t)$ to deviate at all from the solution of (\ref{eq:Friedmann})?  
We can address this in the context of Newtonian gravity.  This is relevant because
Newtonian gravity should provide a very accurate description of the local
universe since all velocities -- Hubble and peculiar -- are small.
And it is in the relatively local universe that the current expansion
rate -- a problem for matter dominated cosmology in the conventional
framework -- is measured. As we shall discuss in more detail below,
at $z < 0.1$ for example, where $H_0$ is reliably measured, the
background can be treated as Newtonian to a precision of order
$z^2 \simeq 0.01$ and corrections to lowest order weak field
gravity perturbations are suppressed by at least a factor $v_{\rm pec} / c$, the ratio of peculiar velocities
to the speed of light.   Also, the absolute value of the curvature
radius, which is arguably a non-Newtonian construct and which
may be identified with $a$, is not relevant here.
All that counts is the expansion
rate $\dot a / a$ and how $a(t)$ changes with time.  In a homogeneous 
model these are determined locally.  The question of how inhomogeneity
affects the expansion might seem to be more complex, but it would seem bizarre indeed if the
expansion rate of the local universe were affected by the emergence of
structure in the distant universe.  So if backreaction is at all important it
should be revealed in a Newtonian analysis.

We will now show that, despite the apparently questionable assumption
of homogeneity in (\ref{eq:Friedmann}), the system of equations 
(\ref{eq:Friedmann}-\ref{eq:poisson}) is actually precisely
equivalent  to the full Newtonian equations of motion.  

\section{Newtonian Cosmology in Scaled Coordinates}

For $N$ particles of mass $m$ interacting
under their mutual gravitational attraction there are $3N$ second
order differential equations
\begin{equation}
\ddot \br_i = G m \sum\limits_{j \ne i} \frac{\br_j - \br_i}{| \br_j - \br_i|^3} .
\label{eq:ddotr}
\end{equation}
These may be solved numerically provided initial positions $\br_i$ and
velocities $\dot \br_i$ for the particles.

Writing this in terms of arbitrarily re-scaled coordinates
$\br = a(t) \bx$, so $\dot \br = \dot a \bx + a \dot \bx$ and $\ddot \br = \ddot a \bx + 2 \dot a \dot \bx + a \ddot \bx$, (\ref{eq:ddotr}) 
becomes
\begin{equation}
\ddot \bx_i + 2 \frac{\dot a}{a} \dot \bx_i= \frac{G m}{a^3} \sum\limits_{j \ne i} \frac{\bx_j - \bx_i}{| \bx_j - \bx_i|^3}
- \frac{\ddot a}{a} \bx_i .
\label{eq:ddotx1}
\end{equation}

What we are interested in is the motion of particles
with initial conditions that are close to being in uniform Hubble expansion
with some initial expansion rate $H$ (very close if we start at early times).  So we might lay down
particles on a regular grid in $\br$-space within some very  large spherical
boundary centred on the origin and give the particles small displacements $\delta \br$
and velocities $\dot \br = H \br + \delta \dot \br$
with `peculiar' velocities $\delta \dot \br$ chosen to excite the growing mode.
The corresponding initial conditions in terms of $\bx$-coordinates are
\begin{equation}
\bx = \br / a \quad {\rm and} \quad \dot \bx = ((H - \dot a / a)\br + \delta \dot \br) / a.
\label{eq:ics}
\end{equation}

The sum in (\ref{eq:ddotx1}) will have two components:  A `zeroth order' acceleration
that, in the limit that the grid spacing becomes very small, 
is the same as the gravitational acceleration of a uniform density sphere,
which grows linearly with $\bx_i$, plus a perturbation determined by the
displacements from the grid (we may think of the source of the gravity being
that of the unperturbed grid of particles plus that of a set of dipole sources).
If we define the number density of particles in $\bx$-space $n(\bx) \equiv \sum_i \delta(\bx - \bx_i)$
and $\delta n \equiv n - {\overline n}$ with ${\overline n}$ the inverse of the
grid cell volume in $\bx$-space, equations (\ref{eq:ddotx1}) become
\begin{equation}
\begin{split}
\ddot \bx_i + 2 \frac{\dot a}{a} \dot \bx_i - & \frac{G m}{a^3}  \int d^3 x\; \delta n(\bx)  \frac{\bx - \bx_i}{| \bx- \bx_i|^3}\\
& = - \left(\frac{\ddot a}{a} + \frac{4 \pi G m {\overline n}}{3a^3}\right) \bx_i .
\end{split}
\label{eq:ddotx}
\end{equation}

It is interesting to compare this with the conventional equations.  Those are $3N + 1$ equations
(3 per particle plus the Friedmann equation for $a$) whereas here we have only $3N$
equations, just as in (\ref{eq:ddotr}).   

But since $a(t)$ is arbitrary we may {\em assert\/} that $a(t)$ is such that the
RHS of (\ref{eq:ddotx}) vanishes -- i.e.\ that $a(t)$ is a solution of (\ref{eq:Friedmann}) --
in which case the vanishing of the LHS is equivalent to the conventional structure
equations (\ref{eq:vdot}) and (\ref{eq:poisson}).  

Moreover, if we set the initial conditions for (\ref{eq:Friedmann}) to be
$\dot a / a = H$ then we see from the second of (\ref{eq:ics}) that
$\dot \bx = \delta \dot \br / a$; the initial velocity in $\bx$-space is a pure perturbation
with no Hubble-flow component.

Alternatively, if one does not require (\ref{eq:Friedmann})
one obtains modified `structure' equations with a large-scale radial acceleration that would drive a
Hubble-like flow to compensate. The results for all
physical quantities such as positions, velocities, density etc.\ however are all invariant
with respect to the choice of $a(t)$.

We thus obtain the original conventional system of equations,
in which there is no feedback (or `backreaction') from the structure
equations on the expansion.  
But this is no longer open to the challenge that
(\ref{eq:Friedmann}) is only an approximation.  Equation (\ref{eq:ddotx})
is precisely equivalent to (\ref{eq:ddotr}); we are simply using the freedom 
in choice of $a(t)$ to impose (\ref{eq:Friedmann}) as an identity. We emphasise
that the resulting system of equations -- the basis of `Newtonian cosmology' -- is not novel.  
It was first obtained by Dmitriev \& Zeldovich (1964)
and is what is integrated in essentially all modern N-body simulations.  Equivalent
equations of motion were also obtained by Peebles (1989)
in the context of reconstruction of local group orbits from 
the action principle.  The difference here is mainly one of perspective.  We have shown 
that, in principle, the scale factor is arbitrary
and need not obey  (\ref{eq:Friedmann}) but, in that case, one must then also modify
the `structure' equations accordingly.  We note that Newtonian cosmology
with point mass particles was also considered by Ellis \& Gibbons (2015) 
who considered a model in which there a population of `background' particles 
with no peculiar motions and `galaxies' that respond to their own peculiar
gravity and the mean-field gravity of the background particles.

As discussed by Dmitriev and Zel'dovich (1964), Newtonian cosmology is
obtained by considering perturbations to a large uniform density expanding
sphere.  The radius $R$ of this sphere may be taken to infinity within Newtonian
physics as all the physically observable quantities are regular in that limit.
In that sense the results are insensitive to the `boundary conditions
at infinity'. It is however required that one consider a sphere as any other
geometry would not expand isotropically and homogeneously.  Within the
infinite sphere the structure equations (\ref{eq:ddotx}) may be used to describe
structure that is periodic within some finite size box of side $L$, in which
case the peculiar potential, velocity and displacements may be expanded
as Fourier series as is commonly done.

\section{Discussion}

We have tried to clarify the meaning of the conventional
equations of Newtonian cosmology.  We have expressed the
usual Newtonian equations (\ref{eq:ddotr}) in terms of
re-scaled (or what cosmologists call `comoving') coordinates $\bx$ to obtain (\ref{eq:ddotx}).
But in these equations the scale factor $a(t)$ is completely arbitrary and has no physical
impact so there is no dynamical equation that $a(t)$ must obey.
This reflects the fact that the universe we live in can, if one so wishes, be considered
to be a perturbation of some hypothetical `background' cosmology,
but there is freedom in choosing the background.  
Exploiting this freedom, the Friedmann equation (\ref{eq:Friedmann}) may
be asserted as an identity, and with the initial conditions set to
$\dot a/ a = H_0$ we have shown that we then obtain the conventional
equations of cosmological dynamics.  In these equations (\ref{eq:Friedmann}) should
not be considered a dynamic equation so much as an auxiliary relation
that determines the form of the equations of motion of the particles.

Newtonian dynamics does not strictly {\em require\/} that the scale factor
obey the conventional Friedmann equation. 
But if $a(t)$ is chosen {\em not\/} to obey the Friedmann equation
this results in an additional long-range radial  force
in the equations of motion in $\bx$-coordinates;
the RHS of (\ref{eq:ddotx}).  This is required in order that physical 
quantities like the expansion
rate be independent of the choice of scale factor.
Similarly, if the initial $\dot a / a$ is not taken to
agree with the initial physical expansion rate this implies
initial conditions where there will be a net expansion or
contraction in comoving coordinates. 
So if (\ref{eq:Friedmann}) is violated, or the initial conditions
are not set appropriately, the solutions of the `structure' equations
no longer just describe the emerging structure; they also
include part or all of the `background' evolution.

The fully non-linear dynamics of the local universe are exactly 
described using the standard equations in the Newtonian limit.
In these the evolution of the scale factor is decoupled from the
evolution of structure, and is fixed by the initial density
and expansion rate and, of course, the presence of dark energy. 
There is no Newtonian backreaction on $a(t)$ from structure.

Specifically, one cannot, as Racz et al.\ have proposed,
keep (\ref{eq:vdot}) and (\ref{eq:poisson}) but modify
(\ref{eq:Friedmann}).  These equations are seen from
(\ref{eq:ddotx}) to be intimately linked together.  To modify
(\ref{eq:Friedmann}) alone results in a theory that does not reduce to Newtonian gravity in the limit
of small velocities as does Einstein's gravity.

To remind ourselves why this is important, 
this means that a matter only universe, with baryon and dark matter
densities (in relation to radiation density) set at values that are
acceptable for big-bang nucleosynthesis and CMB acoustic peaks,
cannot be successfully matched to observations.  
As is well known, if the density parameter is taken  to be unity
this will result in an unacceptably small final expansion rate and
if a low $\Omega$ is chosen this would result in global
hyperbolic spatial curvature that would mess up the angular
scale of the CMB ripples.

How does this relate to the `generalised Friedmann equation' 
(\ref{eq:BE97}) of  Buchert \& Ehlers (1997)?
It is important to realise that their formula has a very different
meaning to the Friedmann equation that appears
with the structure equation in the conventional framework.
Their $a$ is the cube root of a particular volume $V$ and their
equation describes the relationship between $\ddot a / a$
and the density within that volume.
It is not at all surprising that the  $\ddot a / a$ for some
particular volume would differ at some level from
$- 4 \pi G M / 3 a^3$ if there is inhomogeneity.
The acceleration is some combination of the background plus
fluctuation and the mass density is similarly the background
density plus the density fluctuation.  But these two fluctuations
need not be the same.  Indeed, it is perhaps surprising
that the deviations would appear only at second order
in the fluctuations and not be already present in linear theory.
But one would hardly call this `backreaction' of structure
on the global expansion rate; it is simply inhomogeneity
affecting the local expansion rate and local density but
in slightly different ways.  The key question is really whether
there is a {\em systematic\/} difference.
If the combination of quantities
being volume averaged in ${\cal Q}$ has a non-zero expectation value then
this would imply deviations from Friedmann behaviour
even in the limit that $V \rightarrow \infty$ and one would
have to reject (\ref{eq:Friedmann}) in favour of (\ref{eq:BE97}).

But this is not the case.  A strong indication of this, as
shown by Buchert \& Ehlers, is that ${\cal Q}$ can also be expressed as a surface
integral.  They obtained this by 
decomposing  the total velocity into a Hubble flow plus perturbation
$\bv = H \br + \bu$ with the expansion rate being that of the 
region in question.  More relevant is to consider the peculiar
velocity with respect to the global expansion rate.  As shown in the appendix, this gives
\begin{equation}
{\cal Q} =  \frac{1}{V} \int {\bf dA} \cdot (\bu (\bnabla \cdot \bu) - (\bu \cdot \bnabla) \bu) 
- \frac{3}{2V^2} \left[\int  {\bf dA} \cdot \bu\right]^2
\label{eq:Qsurfint}
\end{equation}
where the first term is equation 14 of  Buchert \& Ehlers and the second term appears
in their appendix B.

An obvious, but largely unanswered, question is: How does $\cal Q$ in (\ref{eq:Qsurfint}) depend on $V$?  And how large is it typically?
An under-appreciated feature of (\ref{eq:Qsurfint}) is that, as discussed in the appendix, 
the expectation value of the integrand of the first term vanishes by symmetry.  
Consequently, the average of this term, taken over an ensemble of volumes of any size, also vanishes.
The typical value of the fluctuation in this contribution to $\cal Q$ for a volume of size $r$ is $|{\cal Q}| \sim v^2 / r^2$,
independent of the `coherence length' $\lambda$ of the peculiar velocity field.  This tends to zero as $r \rightarrow \infty$,
and should be considered to be a `cosmic variance' fluctuation.
The second term has a non-zero expectation value, but this is of order $\langle {\cal Q} \rangle \sim v^2 \lambda^2 / r^4$
and falls to zero even faster.

Thus the quantitative answer to the question that Ellis posed and Buchert \& Ehlers addressed is that, averaged over
large volumes, the scale factor does obey the Friedmann equation and there
is no backreaction on $a(t)$ from the emergence of structure, consistent
with what we have found above. 

t is reasonable to ask how, if at all, the conclusions here differ from the 
current position
of experts in the backreaction community.  In the first paragraph
of Buchert \& R\"as\"anen (2012) they say that `In standard linear
theory, the effect vanishes on average by construction. In Newtonian
gravity, this turns out to be true also in the non-perturbative
regime.'  This is not in conflict with what we have found here.
However, a key phrase here is `by construction'.  
Expanding on this they say `When we impose periodic
boundary conditions in Euclidean space, the backreaction variable
$\cal Q$ is strictly zero on the periodicity scale (a three-torus has no
boundary)'.  Similarly, Buchert, Kerscher \& Sicka (2000) say
`Note that both the numerical and analytic approaches {\em enforce\/} a
globally vanishing backreaction by imposing periodic boundary
conditions'. This connection between vanishing of $\cal Q$ and periodic
BCs is repeated, and later, in their discussion of N-body simulations one
reads that  `Most cosmological NÐbody
simulations solve [....] with periodic boundary conditions. Hence,
the boundary of $C$ is empty [....], and from Eq. (10) we directly
obtain ${\cal Q}_C = 0$.' and, following this, `It will be a challenge to
incorporate backreaction effects in NÐbody simulations.'  

We think it might be possible for a reader 
of these papers to come away with the impression
that the N-body simulations and analytic calculations
are missing some extra non-negligible backreaction physics `by construction'
through the special choice of periodic boundary conditions.
This might be further reinforced by Buchert \& Ehlers statement that
for $\cal Q$ to be zero would require a `conspiracy'.

What we have shown here that is $\cal Q$ tends to zero
very rapidly in the limit of large volumes regardless
of whether the structure is assumed to be periodic. This
is based solely on the symmetry properties of
statistically homogeneous and isotropic velocity fields.
Another minor novelty of our approach is to show how the surface integral
form for $\cal Q$ may be obtained directly rather than through
the intermediary step of Raychaudhuri's equation.

The analysis leading to our (\ref{eq:ddotx}) represents a
significant advance over the approach followed in e.g.\ 
Buchert \& Ehlers and later backreaction studies where it is assumed that
the matter can be modelled as a pressure-free fluid.
Uncondensed baryonic gas may, if the cooling time
is sufficiently short, approximate such a fluid. Collisionless
dark matter at very early times before non-linear structures form
may behave a lot like such a fluid. But in the non-linear regime
that is relevant here this assumption is, at the very least, highly
questionable.  Once multi-streaming
occurs, collisionless dark matter and galaxies
develop pressure.  The same is true for the bulk of the
baryonic gas which cannot cool efficiently. It is only in this
way that realistic equilibrated (i.e.\ `virialised') or quasi-equilibrated
structures can form.  The only equilibrium state for a pressure-free
gas is, in contrast, a dense rotationally supported disk.

The analysis here has been entirely Newtonian.
It is certainly true that there must be genuine relativistic effects that
will modify the expansion rate.  One such effect
is that of intergalactic pressure.  It is known that most galaxies
harbour black holes and it is thought that these merge in the
process of the merging of their hosts.    The rapidly time varying 
gravitational potential will inevitably result in expulsion of a small
amount of stars and dark matter at high velocities.   This results in non-zero
kinematic pressure in intergalactic space which, owing to the expansion,
will do $P dV$ work.  According to special relativity, $\delta E = \delta mc^2$, so this loss
of energy results in a decrease in mass and therefore a modification to the continuity equation; i.e.\ there
will be a non-zero, and positive, pressure $P$ in $\dot \rho = - 3 H (\rho + P / c^2)$.  
This pressure will also appear in the Friedmann $\ddot a$ equation and will result in a modification to
the expansion rate.  Simple estimates, however, suggest
that this is negligible for all practical purposes.  

One might naively
question whether the pressure inside bound stellar systems or in
stars themselves might need to be included in some average sense
in the Friedmann equations.  That is not the case, as was shown by
Einstein \& Straus (1945).  The title of their paper was `The Influence of the Expansion of Space
on the Gravitation Fields Surrounding the Individual Stars' and they
concluded that there is none.  The fact that distant matter is expanding
away from stars does not affect them; their gravitational mass -- the parameter defining the
Schwartzschild geometry that surrounds them -- is fixed.  Consequently
the gravitational mass density of a population of stars, black holes or other compact
objects must dilute as $1/a^3$ so the pressure $P$ in the continuity
equation (and consequently also in the acceleration equation) must
vanish. 

It has been proposed (e.g.\ Buchert \& R\"as\"anen 2012) that there may be strong 
GR backreaction on the expansion.
We would argue that something quite radical is required for this to be the case.  Imagine that we
live in an `island universe' much like ours, but extending only to say about 
$z = 0.1$, thus including the region where
the expansion rate $H_0$ is reliably established to be approximately 70 km/s/Mpc.
For a homogeneous sphere, the errors incurred in the Newtonian
approximation -- the difference between the proper and gravitational
mass for instance -- is on the order of $v^2 / c^2$ or about one percent.

Adding structure within the island excites the usual scalar metric
perturbation with only one spatial degree of freedom as this is driven by the density.
Beyond this lowest order weak field are `gravitomagnetic' effects, driven
by the matter 3-current associated with structure, but the metric perturbations
arising from motions are suppressed relative to the Newtonian term by
a factor $v_{\rm pec} / c$.  That such post-Newtonian effects are small is supported
by direct weak-field calculations of Adamek et al (2013).
Beyond the 4 degrees of freedom associated with the matter 4-current,
all that is left are the two metric degrees of freedom of gravitational
waves, but these do not affect the expansion rate as they are traceless.
The errors involved in modelling the expansion of such an island universe with Newtonian
physics should therefore be very small.

What then is the effect of adding the external universe?  If this is
spatially homogeneous and isotropic then, as Einstein \& Straus showed,
there is no effect.  The challenge for backreaction proponents is
to explain how the emergence of {\em structure\/} at great distances can
affect the local dynamics and make any appreciable changes to
the local expansion rate (and thus e.g.\ reconcile the large observed
$H_0$ with that expected in a flat universe without dark energy). 
There are local tidal influences from distant structures, but these
are small and, like gravitational waves, do not affect the expansion.
The problem with believing that this occurs in GR is that a cornerstone
of the theory is that spacetime is locally flat. This means that in the local
universe it is the local matter that controls the dynamics
through the 1st law of thermodynamics (energy conservation), expressed
in the Friedmann continuity equation, and the
conservation of momentum expressed in the Friedmann acceleration equation.

Finally, and returning to Newtonian dynamics, we mention another
probably small but not obviously vanishing cause of backreaction; that
of `tidal torques'.  It is well known that, in conventional models for
galaxy formation, galaxies acquire their angular momentum through
non-linear effects as they depart from the linear regime but before
they decouple.  This can be thought of as a kind of `mode-mode' coupling
between the galaxy scale fluctuations and a larger-scale motion; the
global expansion.  It does not seem entirely obvious that
this has vanishing effect on the expansion of the universe.  But our
main result here shows that, to the extent that the structure is
a statistically homogeneous and isotropic random process, there
can be no such effect.

\section{Acknowledgements}

I am grateful to Kevin Croker, John Learned and Istvan Szapudi for
stimulating discussions on this topic.  It is also a pleasure to acknowledge useful
feedback from Pierre Fleury, Thomas Buchert, George Ellis, Syksy R\"as\"anen,
an anonymous referee and, in  particular, Anthony Challinor.

\appendix 

\section{Acceleration in terms of surface averages}

Raychaudhuri's equation leads to the Freidmann-like (\ref{eq:BE97})
containing the additional term ${\cal Q}$ that is a volume average of 
quadratic scalars constructed from the velocity shear tensor.  Decomposing the velocity into a Hubble-flow plus
perturbation, Buchert \& Ehlers obtained a surface integral expression for
${\cal Q}$. Here we show how this may
be obtained directly.

The rate of change of the volume at some time $t$ is 
$\dot V = \int {\bf dA} \cdot \bv$ where $\bv$ is the velocity and 
${\bf dA} = \bn dA$ is an outward directed surface area element.  
At some slightly later time $t' = t + \delta t$ the rate of change of the volume will be
$\dot V' = \int {\bf dA}' \cdot \bv' = \int dA' \bn' \cdot \bv'$.

The ratio $dA'/dA$ is the determinant of the 2D matrix describing the
mapping from positions in the initial surface to the final surface.
Since $\br' = \br + \bv \delta t$ this is easily found to be $dA' / dA = 1 + (v_{xx} + v_{yy}) \delta t$
to first order in $\delta t$ where we have erected coordinates so the $z$-axis
is parallel to $\bn$ and where $v_{xx} \equiv \partial v_x / \partial x$ etc.  
Similarly, the unit normal changes if $v_z$ varies across the
area element: $\bn \rightarrow \bn' = \bn - (\hat {\bf x} v_{zx} +  \hat {\bf y} v_{zy})\delta t$.
Thus
\begin{equation}
\begin{split}
{\bf dA}' \cdot \bv' & = dA (1 + (v_{xx} + v_{yy}) \delta t) \\
& \times (\bn - (\hat {\bf x} v_{zx} +  \hat {\bf y} v_{zy})\delta t) \cdot (\bv + \dot \bv \delta t)
\end{split}
\end{equation}
and therefore 
\begin{equation}
\begin{split}
d ({\bf dA}  \cdot \bv)& /dt  = ({\bf dA}'\cdot\bv' - {\bf dA}\cdot \bv) / \delta t \\
& =  [\dot v_z + v_z v_{xx} + v_z v_{yy} - v_x v_{zx} - v_y v_{zy}] dA .
\end{split}
\end{equation}
The coordinate frame independent expression of this is easily found by noting
that, in this frame, this is the same as ${\bf dA} \cdot (\dot \bv + \bv (\bnabla \cdot \bv) - (\bv \cdot \bnabla) \bv)$.

The second time derivative of the volume $\ddot V = (\dot V' - \dot V) / \delta t$ is therefore
\begin{equation}
\ddot V  = \int {\bf dA} \cdot [\dot \bv + \bv (\bnabla \cdot \bv) - (\bv \cdot \bnabla) \bv].
\end{equation}
Gauss's law tells us the first term is $\int {\bf dA} \cdot \dot \bv = - 4 \pi G M$.  

To convert this into an expression involving $a = V^{1/3}$ we use 
$3 \ddot a / a = \ddot V / V - (2/3) (\dot V / V)^2$ to obtain (\ref{eq:BE97}) -- i.e.\ equation 9 of Buchert \& Ehlers --
but now with
\begin{equation}
{\cal Q} = \frac{1}{V} \int {\bf dA} \cdot (\bv (\bnabla \cdot \bv) - (\bv \cdot \bnabla) \bv) - \frac{2}{3V^2}  \left[\int  {\bf dA} \cdot \bv\right]^2
\end{equation}

Writing the velocity as $\bv = H \br + \bu$ we find
\begin{equation}
\begin{split}
{\cal Q} = & \frac{H}{V} \int {\bf dA} \cdot (\br (\bnabla \cdot \bu) - (\br \cdot \bnabla) \bu - 2 \bu) \\
& + \frac{1}{V} \int {\bf dA} \cdot (\bu (\bnabla \cdot \bu) - (\bu \cdot \bnabla) \bu) \\
& - \frac{2}{3V^2} \left[\int  {\bf dA} \cdot \bu\right]^2
\label{eq:Qsurface}
\end{split}
\end{equation}
The integrand in the first line is $\bnabla \times (\br \times \bu)$, so its integral vanishes.
The second line is identical to equation 14 of Buchert \& Ehlers. That was obtained
assuming that $3H$ is the volume average of the volume expansion rate within
the particular volume considered and, as they discuss, the last term here, which
appears in their appendix B, enters if $H$ is taken to be the global expansion
rate.

An under-appreciated feature of the surface integral term is 
that the expectation value of the integrand vanishes by symmetry.  This is because it involves products
of the velocity and its spatial derivative like $\langle v_x v_{zx} \rangle$.  This is the
derivative with respect to lag $\br'$, at $\br' = 0$,
of the correlation function $\langle v_x(\br) v_z(\br + \br') \rangle$.  But
for a velocity field that is statistically homogeneous and isotropic -- no further
assumptions are required -- these functions are even functions of $\br'$ so   
$\langle v_x v_{zx} \rangle$ vanishes (Monin \& Iaglom, 1975; see also Gorski 1988).

For an individual volume the contribution to ${\cal Q}$ will not vanish.
If the velocity field has coherence length $\lambda$ the integrand is on the order of $u^2 / \lambda$.
The mean square is $\langle {\cal Q}^2 \rangle \sim N (\Delta A u^2 / \lambda)^2 / V^2$ where
$\Delta A \sim \lambda^2$ and $N = A / \Delta A$.  It follows that the typical contribution 
is $|{\cal Q}| \sim u^2 / r^2$, independent of
$\lambda$.  This becomes very small for large volumes.

The last line in (\ref{eq:Qsurface}), also of 2nd order in $\bu$, differs from the second in that 
it has a non-vanishing (negative) expectation value.
But it is on the order of ${\cal Q} \sim u^2 \lambda^2 / r^4$
and so is even smaller than the second term for large $V$.
We believe this is what  Kazimierczak (2016) has measured in N-body simulations.

\end{document}